\documentclass[preprint,journal]{vgtc}       

\ifpdf
  \pdfoutput=1\relax                   
  \usepackage{graphicx}                
  \DeclareGraphicsExtensions{.pdf,.png,.jpg,.jpeg} 
\else
  \usepackage{graphicx}                
  \DeclareGraphicsExtensions{.eps}     
\fi%

\graphicspath{{figures/}{pictures/}{images/}{./}} 

\usepackage{microtype}                 
\PassOptionsToPackage{warn}{textcomp}  
\usepackage{textcomp}                  
\usepackage{mathptmx}                  
\usepackage{times}                     
\usepackage{cite}                      
\usepackage{tabu}                      
\usepackage{booktabs}                  

\usepackage{enumitem}
\usepackage{tabularx}
\usepackage{multirow}
\usepackage{color}
\usepackage[table, dvipsnames]{xcolor}
\newcommand{\edit}[1]{{\color{black} #1}}
\definecolor{brickred}{rgb}{0.8, 0.1, 0.1}

\onlineid{1381}

\vgtccategory{Research}
\vgtcpapertype{Please Specify}

\title{Staged Animation Strategies for Online Dynamic Networks}

\author{Tarik Crnovrsanin, Shilpika, Senthil Chandrasegaran, and Kwan-Liu Ma}
\authorfooter{
  \item Tarik Crnovrsanin, Shilpika, Senthil Chandrasegaran, and Kwan-Liu Ma are with University of California, Davis. E-mail:\{tecrnovr, fshilpika, schandrasegaran, klma\}@ucdavis.edu
}

\shortauthortitle{Staged Animation Strategies for Online Dynamic Networks}

\abstract{
  Dynamic networks---networks that change over time---can be categorized
into two types: offline dynamic networks, where all states of the
network are known, and online dynamic networks, where only the past
states of the network are known.
Research on staging animated transitions in dynamic networks has focused more on
offline data, where rendering strategies can take into
account past and future states of the network.
Rendering online dynamic networks is a more challenging problem since
it requires a balance between timeliness for monitoring tasks---so that the animations do not lag too far behind the events---and clarity for comprehension tasks---to minimize simultaneous changes that may be difficult to follow.
To illustrate the challenges placed by these requirements, we explore
three strategies to stage animations for online dynamic networks:
time-based, event-based, and a new hybrid approach that we introduce by
combining the advantages of the first two.
We illustrate the advantages and disadvantages of each strategy in
representing low- and high-throughput data and conduct a user study
involving monitoring and comprehension of dynamic networks.
We also conduct a follow-up, think-aloud study combining monitoring and comprehension with experts in dynamic network visualization.
Our findings show that animation staging strategies that emphasize comprehension do better for participant response times and accuracy.
However, the notion of ``comprehension'' is not always clear when it comes to complex changes in highly dynamic networks, requiring some iteration in staging that the hybrid approach affords.
Based on our results, we make recommendations for balancing
event-based and time-based parameters for our hybrid approach.

}


\CCScatlist{ 
 \CCScat{K.6.1}{Management of Computing and Information Systems}%
{Project and People Management}{Life Cycle};
}

\vgtcinsertpkg

\begin{document}

\firstsection{Introduction}

\maketitle

Within the domain of network analysis and visualization, there is now a
growing interest in visualizing dynamic networks: networks that change
over time.
Visualizing dynamic networks is challenging, as acknowledged
by Moody and McFarland~\cite{Moody2005dynamic}, who point out
\textit{``To effectively display the relational structure of a social
network, at least two dimensions are needed to represent proximity, and
that leaves no effective space (on a printed page) to represent time''}.
Animation has since been widely recognized as a viable option
to show changes to networks that occur over time.
However, most techniques introduced to visualize temporal changes in a
network are demonstrated on \textit{offline} dynamic networks, where
data is collected and processed offline.
Such techniques take advantage of foreknowledge of changes to the
network to scale their representations and/or stage any animations
appropriately.

On the other hand, visualizing online dynamic network data is
challenging since it is generally not possible to predict the future
state of the network, or the rate at which new data is generated.
\edit{
It may not even be feasible to identify \textit{what} behavior to
look for, especially if it is previously unseen.
Once the behavior is characterized unambiguously, algorithms can be
devised to recognize it, but in mission-critical applications,
visual display observed by humans is the most practical option to
recognize new, unexpected patterns.
For instance, security personnel monitor CCTV footage in real time,
looking for indicators of threatening events.
These ``indicators'' cannot always be coded to be recognized by an
automated system; it depends on the experience and perceptiveness of the
person monitoring the footage.
}

\edit{
In network visualization, an analogy can be drawn from 
computer security monitoring, for instance network flow events within
the Los Alamos National Laboratory's (LANL) corporate, internal computer
network~\cite{Kent2015comprehensive}.
These events indicate network connections between computers and contain
relevant information such as time of connection, duration of connection,
amount of information moved, and the protocol used.
Authorized attackers testing network security often create ``compromise
events'' as part of their tests~\cite{akent-2015}.
These events are recorded with information such as \textit{domain},
\textit{source computer}, and \textit{destination computer}.
Models are yet to be created to automatically identify some of the
compromise events that are recorded in this dataset\cite{akent-2015}.
Not all malicious behavior recorded in the dataset have validated or
correlated indicators.
Such events thus need to first be observed and understood manually.
Depending on the data sensitivity, real-time visual monitoring could supplement
other automated tracking approaches.
}

Online dynamic network data can be discrete or continuous; i.e.,  changes
to the network can occur either in fits and spurts, or gradually over
time~\cite{Moody2005dynamic}.
A similar variation can occur with the rate of incoming data as well;
i.e., in some instances, a large number of entities (or relations) can
appear/disappear over a short time interval, while in other instances
there is very little incoming data on entities/relations.
\edit{A real-time animation strategy can represent these changes with high
fidelity---events are shown almost as soon as they occur---but the rate
the strategy may fail in providing the viewer with useful
information if there are too many changes for the viewer to track.}

Alternatively, a staged animation strategy may be devised to keep the
user informed of significant changes to the network by ``binning''
changes until a threshold of time or changes is reached before animating
them.
However, this strategy has its potential shortcomings: a purely
time-based binning may suffer from the same drawback as the real-time
strategy when too many changes occur within the fixed time interval.
A purely event-based staging strategy addresses the comprehension issue
by ensuring that the user does not get overwhelmed by changes to the
network in a single stage.
This could mean that the animation staging and the duration of the animation for each stage could cause a bottleneck, increasing the time lag between the event occurring and when it is shown to the user.
In addition, this could also result in long wait times between updates
for low data rates.
Staged animation strategies have the added requirement of devising
means to inform the users about the cause behind re-organizations
to the network~\cite{Bach2014graphdiaries}, which may be implicitly
apparent in a real-time strategy.
Any animation technique developed to visualize online dynamic networks
thus needs to address both requirements: provide the viewer with timely
and understandable ``chunks'' of information on both \textit{what}
changes have occurred \edit{to the network and/or layout} and \textit{why}, without significantly
compromising the time lag between events and their on-screen depictions.

In this paper, we compare time-based and event-based staging strategies,
along with an alternative, ``hybrid'' staging strategy that we
introduce.
The hybrid staging strategy uses a combination of time-based and
event-based staging, whichever threshold is reached earlier.
We motivate each strategy by examining the requirements of animations
when it comes to monitoring and comprehending dynamic networks.
We evaluate all three strategies through a within-subjects study with 21
participants who perform a series of monitoring and comprehension tasks
with each animation staging strategy.
A comparison of task performance for each staging strategy shows that time-based staging strategies lower the time lag between an event and its depiction in the animation.
However, users prefer and perform better with event-based staging strategies as they allow for better comprehension.
Hybrid staging performs equivalently to event-based staging for most of
the tasks.
These findings are further validated by a think-aloud qualitative study of the animation strategies with network visualization experts.
The contributions of our work include
(1) a design space of animation
strategies meant for representing changes to a network in real time,
(2) an evaluation of the strategies based on representative tasks relevant
to real-time network comprehension and analysis, and
(3) an examination of the limit of the hybrid strategy through a simulation of varying event occurrence rates over varying time intervals with all three animation strategies.
Based on our findings, we provide alternative directions to explore for hybrid staging strategies.

\vspace{-0.10in}

\section{Related Work}
A comprehensive review of dynamic network visualization techniques is
provided by Beck et al.~\cite{Beck2014}, who categorize the techniques
into \textit{animation} or ``time-to-time mapping'', \textit{timeline}
or ``time-to-space mapping'', and hybrids of both approaches.
In this section, we will briefly cover all three techniques with a
greater emphasis on time-to-time mapping given the scope of our paper.
We will also examine differences between offline and online dynamic
networks and the challenges presented by the latter, and go over some
basic principles of animation that inform our exploration of the design
space we consider for our study.

\subsection{Static Representations of Dynamic Networks}
\label{sec:dynamic}

The challenge in visualizing dynamic networks lies in accurately
depicting the changes that occur to the network over time, while also
ensuring that salient changes are perceptually more apparent to the
user.
Timeline-based approaches achieve this by providing an overview of
changes that occur between time steps.
For instance, TimeArcTrees, a technique introduced by Greilich et
al.~\cite{Greilich2009} vertically aligns graph nodes across time steps,
facilitating comparison between two states based on position of the
nodes and the organization of edges between nodes.
While graph scale is somewhat addressed by collapsing subgraphs into
parent nodes, comparison between two graphs still requires a one-to-one
comparison between nodes.
Burch et al.~\cite{Burch2011} address this issue in TimeArcTrees by
introducing parallel edge splatting: each time-state of the graph is
first mapped to a 1D vertical layout, with edge overplotting and
weighting shown as a heatmap between parallel lines. 
Reda et al.~\cite{Reda2011} address issues of scale by focusing on
communities: each entity in the network is plotted as a polyline
extending from left to right, with vertical movements of the polyline
corresponding to its membership in the communities that exist within the
time period of interest.
Polylines that stay in the same community over time are bundled together
to form a band, further addressing issues of scale.
Vehlow et al.~\cite{Vehlow2015} improve on this work by using Gestalt
principles to show continuity between communities as they merge or split
over time, and use color to depict similarities between communities
over time.
Though timeline-based representations actively focus on the issue of
scale, they are restricted to a finite screen space in which to show all
salient states of the graph.
\edit{Previous works~\cite{simonetto2017drawing, SimonettoEB} have used a model for dynamic graphs which is not based on time slices using the DynNoSlice force-directed algorithm which uses a space-time cube (2D+time) to visualize an offline dynamic graph.}
Static representations of \edit{dynamic graphs}
have been shown to work better than dynamic representations for analysis\edit{~\cite{Archambault2011, Farrugia11, brandesunrolling, simonetto2017drawing, SimonettoEB}} but such representations are more suitable for offline dynamic graphs, where the states of the network are known in advance.
\edit{Real-time data would produce a larger number of static states as each event can be considered a new state without periodical summarising or binning of the states. }

\subsection{Animated Representations of Dynamic Networks}
In contrast, animation-based techniques---typically based on node-link
diagrams---offer a more intuitive visualization of dynamic networks.
Force-directed layout generation is one of the most common approaches
for visualizing networks.
The method produces aesthetically-appealing layouts for static networks,
but is also useful in visualizing dynamic networks as changes to the
network---modeled as particles entering/leaving the system and coupling
with/decoupling from other particles in the system---can result in
smooth animations.
However, force-directed approaches are not without fault: minor changes
to the graph, such as the linking of two disconnected components, can
have a large impact on the overall layout.
Because position is used to encode entity relations, a user can form and
maintain an abstract interpretation of the network's structure, called a
``mental map''~\cite{Diehl2002}.
In their review of dynamic graph visualizations, Beck et
al.~\cite{Beck2014} identify the main goal of animation-based techniques
to be the preservation of this mental map, typically by keeping the
position of nodes in node-link diagrams stable over time, thus
minimizing the visual difference between the layouts of the
network across different time slices.

\subsubsection{Animated Transitions in Offline Dynamic Networks}
Most strategies for computing transitions between dynamic network states
have been developed for ``offline'' dynamic graphs, whose structure over
time is known at the time of visualization~\cite{Beck2014}.
Computing strategies thus use this ability to ``look ahead'' and
anticipate changes that inform the current layout.
For instance, Diehl and G{\"o}rg~\cite{Diehl2002} introduced a metric
called ``mental distance'' to indicate the similarity between two
layouts, and built a metagraph from the time sequence to help preserve
the mental map.
They also introduced a \textit{Foresighted Layout with Tolerance} (FLT)
approach for force-directed graphs that trades layout quality of
individual graphs for overall graph stability to preserve the mental
map. 
G{\"o}rg et al.~\cite{Gorg2004} extended FLT to hierarchical and
orthogonal layouts and developed adjustment strategies for each.
Initial approaches to lay out dynamic graphs used the GRIP
algorithm~\cite{Gajer2000grip} for its speed in laying out large graphs.
For instance, 
Collberg et al.~\cite{Collberg2003} added time-slice information to the
GRIP algorithm to compute distances between corresponding nodes from
consecutive time slices of the same graph.
A similar approach was used by Erten et al.~\cite{Erten2003} who
modified GRIP to preserve the mental map by minimizing positional
changes between nodes in one timeslice to the corresponding nodes in
subsequent timeslices.
Brandes et al.~\cite{Brandes2006} argue for the suitability of spectral
layouts to visualize dynamic networks, as layout changes scale
proportionally to changes in the graph.
They demonstrate their argument using spectral methods to animate
small-world network models over time.
More recently, GraphDiaries~\cite{Bach2014graphdiaries} combines the
advantages of timeline and animated approaches to represent dynamic
graphs.
It uses small multiples to provide an overview of the network
states over time, and uses staged animation transition to show the user
what changes between two given states.

When visualizing large dynamic networks, it becomes infeasible to track
individual node positions, and approaches focus instead on tracking
clusters.
Kumar and Garland~\cite{Kumar2006}, for instance, propose a
stratification technique to visualize hierarchies in large graphs, and
extend their approach to dynamic networks by tracking the changes in the
clustering metric over time and minimizing changes in node positions. \edit{A similar technique of clustering nodes that share a common motion from initial layout to the target layout is mentioned in ~\cite{FriedrichGraphinMotion}}.
Sallaberry et al.~\cite{Sallaberry2012} use their previous rapid
graph layout approach~\cite{Muelder2008} to compute clusters in a given
dynamic network at each time step, and use supporting views to visualize
how clusters evolve over time.

\subsubsection{Animated Transitions in Online Dynamic Networks}

The main challenge for online dynamic networks is that visualization
strategies can only take into account the past states of the network as
the future states are unknown.
Unknown future states of a network can result from the data itself, with
the addition or removal of new entities and/or relationships.
They can also result from user interactions, such as filtering or
reorganizing data.
In both cases, the objective remains the same---preserving the user's
mental map of the network---while the challenges are different.
Most prior work on visualizing online dynamic networks address changes
due to user interactions rather than changes due to streaming data.
Early approaches to address the inherent unpredictability of these
networks used random field modeling where models of the layout and its
stability were used by a stochastic estimator that computes layouts in
terms of random fields~\cite{Brandes1997}, but ignored mental maps when
computing the layouts.
Brandes et al.~\cite{Brandes2006} whose spectral layout is discussed in
Section~\ref{sec:dynamic} argue that their approach can be applicable to
online dynamic networks, as it is based on the assumption that since the
graphs in consecutive time steps are similar, spectral layouts of these
graphs do not vary significantly.

However, there has been some work done on online graphs that involve
``streaming'' data.
In early work, North~\cite{North1996} proposed a heuristic
called DynaDAG for directed acyclic graphs to view incremental updates
to graph layouts based on a combination of operational primitives.
Lee et al.~\cite{Lee2006} use simulated annealing to address the
challenge of preserving the mental map while addressing changes in the
network from streaming data, but at the cost of speed as the
approach redraws the entire layout at each time step.
Frishman and Tal~\cite{Frishman2004} introduce ``spacer vertices'' that
minimize the movement of graph clusters between successive layouts as a
technique to preserve the mental map between updates.
In a later work~\cite{Frishman2008}, they address the same problem by
proposing an algorithm that uses a \textit{pinning weight} to identify
computationally-intensive parts of the network, i.e. areas of the
network that change over time, and computes their layouts using the GPU.
Gorochowski et al.~\cite{Gorochowski2012}
suggest a similar approach: an ``age-directed layout'' technique that both colors and
adjusts the degree of a node's movement based on its ``age''.
Nodes that show fewer updates to their connections over time are
considered ``older'' and move less, whereas ``younger'' nodes are
subject to greater movement.
Hayashi et al.~\cite{Hayashi2013} use two techniques to manage changes
to the graph: an automatic edge resizing technique reduces variations in
edge lengths over time, while a sorted sequential barycenter merging
technique updates the graph with new nodes based on their connectivity
to the existing nodes.
Both techniques reduce variation in the graph to preserve the mental map.

In more recent work, Crnovrsanin et al.~\cite{Crnovrsanin2015} use an
incremental algorithm based on $FM^3$~\cite{Godiyal2008}, which uses
GPU acceleration for fast computation.
The incremental algorithm, which enables fast computation between
updates while preserving the mental map, is followed by a refinement
technique that uses an energy-minimization strategy to reduce edge
crossings and lengths for an aesthetic layout.
In our study of staging strategies, we use this incremental
algorithm to preserve the mental map and refine the graph layout between
animation stages.

\subsection{Staged Animation and Perception}
In the prior sections we looked at static and animated
representations of dynamic networks.
While static representations can provide an excellent overview,
animation provides a metaphor for transition (in the form of movement)
consistent with our mental model of the physical world.
In addition, animations can convey not only transitions, but also
causality~\cite{Scholl2000perceptual} (e.g. move nodes apart after deleting an
edge between them).
Animated transitions are widely used in human-computer interaction (HCI)
for providing fluid interactions and in information visualizations to
maintain continuity between different states of
visualizations~\cite{Elmqvist2011fluid}.

While layout strategies optimize the use of space---specifically, the
use of displacement---to preserve mental maps, animations can be said to
use \textit{time} for the same purpose.
However, animation needs to be used with care. 
When animations show complex interactions of moving parts, the viewer's perception of transitions may be inaccurate~\cite{Tversky2002animation}.
Moreover, there are perceptual limits to the number of
independently-moving objects our visual attention can simultaneously
track~\cite{Pylyshyn1988tracking}.
This ability is further impacted by the speed of the animation; faster animations reduce the number of objects we can track
simultaneously, while also reducing the tracking
accuracy~\cite{Feria2013speed}.
Strategies to minimize the number of ``objects'' that the user needs to
track typically discretize objects over time and/or space.
Trajectory bundling, for instance, uses discretization over space, 
``grouping'' individual objects and moving them together, thus
taking advantage of the Gestalt principle of common fate to improve
the viewer's ability to track more points~\cite{Du2015trajectory}.
On the other hand, staged animation~\cite{Heer2007animated} uses the
discretization-over-time approach, where animated transitions are presented in stages over successive time intervals.
While users consistently prefer staged animations over simultaneous
ones, it is not always recommended.
Staging can increase the time span between events, rendering the user's short-term memory unreliable~\cite{Bach2014graphdiaries}.

When it comes to dynamic graph visualization, there is a limit to the
amount of trajectory bundling that can be done while preserving the
mental map of the graph, making staged transition a promising strategy.
GraphDiaries~\cite{Bach2014graphdiaries} is the only work that, to our
knowledge, examines strategies in staging animations in dynamic graph visualization.
They split the staging into element removal, transformation and addition in that order.
This order is shown to be optimal in reducing ambiguity while also
minimizing transition time.
However, GraphDiaries is designed for \textit{offline} dynamic graphs,
allowing for the choice of staging strategies ahead of time to reduce the
user's perceptual load. 
\edit{Wang et al.~\cite{nonuniwangetall} investigate a non-uniform time slicing approach by adapting histogram equalization to create time slices of equal visual complexity, thus conveying more important details of time slices when there is a sudden influx of edges.
However, due to the lack of user studies, it remains unclear which graph analysis task benefits from the non-uniform and uniform time slicing approach.}

In this paper, we explore animation strategies for \textit{online}
dynamic graphs, wherein we adapt staging order and timing from
GraphDiaries but study the effects of time-based and event-based
animations, as well as a hybrid of the two approaches that we introduce.

\section{Design Considerations}\label{sec:design_considerations}
Analyzing changes to dynamic networks in real time involves two main
kinds of tasks: \textit{monitoring}, where the analyst needs to
become aware of a particular kind of change as soon as it occurs, and
\textit{comprehension} where the analyst needs to understand what
changes have occurred to a graph over a period of time.
For real-time analysis, it is safe to assume that this period of time is
relatively short.
From the related research, it is clear that there are several tradeoffs
that need to be made in order for the viewer to keep track of the
changes.
The balance lies between keeping the analyst aware of changes to the
network as soon as they occur, yet not overwhelming them with too
many updates to the network.
We posit that the design of staged animated representations of
change in online dynamic graphs needs to balance three aspects:
\vspace{-2mm}

\begin{enumerate}[itemsep=-0.5mm]
  \item[R1] Timeliness: The animated representation of the event should
    occur as close as possible to the actual time of the event.
  \item[R2] Mental Map preservation: The animated representation should
    occur in a way that allows the viewer to track the changes to the
    graph, thus preserving their mental map of the graph.
  \item[R3] Minimize Transition Time: The animated transition should be
    short enough to make effective use of the viewer's short-term memory
    in recalling the changes in the graph.
\end{enumerate}

\vspace{-2mm}
For the purpose of this paper, we focus on changes to the network that
are triggered by new data, rather than by user interactions.
We use Chevalier et al.'s definitions of transition and
animation~\cite{Chevalier2014not} to define the design space of staged
transitions.
They define \textit{transition} as a pair of visual states (initial and
final).
An \textit{animation} is a series of images that provides the impression
of perceptual continuity between the initial state of the transition and
the final state.
We also define, for the purpose of our study, an \textit{event} as a
change in the network.  An event can thus be (a) the appearance of a new
node, (b) the appearance of a new edge, (c) the disappearance of an
existing node, (d) the disappearance of an existing edge, and (e) a
combination of (a) and (b) or of (c) and (d).
We define a \textit{stage} of the animation as a representation of one
or more events collected based on a triggering parameter.
Since events occur over time, we consider both time
intervals and event count as triggering parameters for these stages.
Finally, we define \textit{animation time} as the duration over which
the animation plays out on the interface.

\begin{figure*}[t]
  \includegraphics[width=\textwidth]{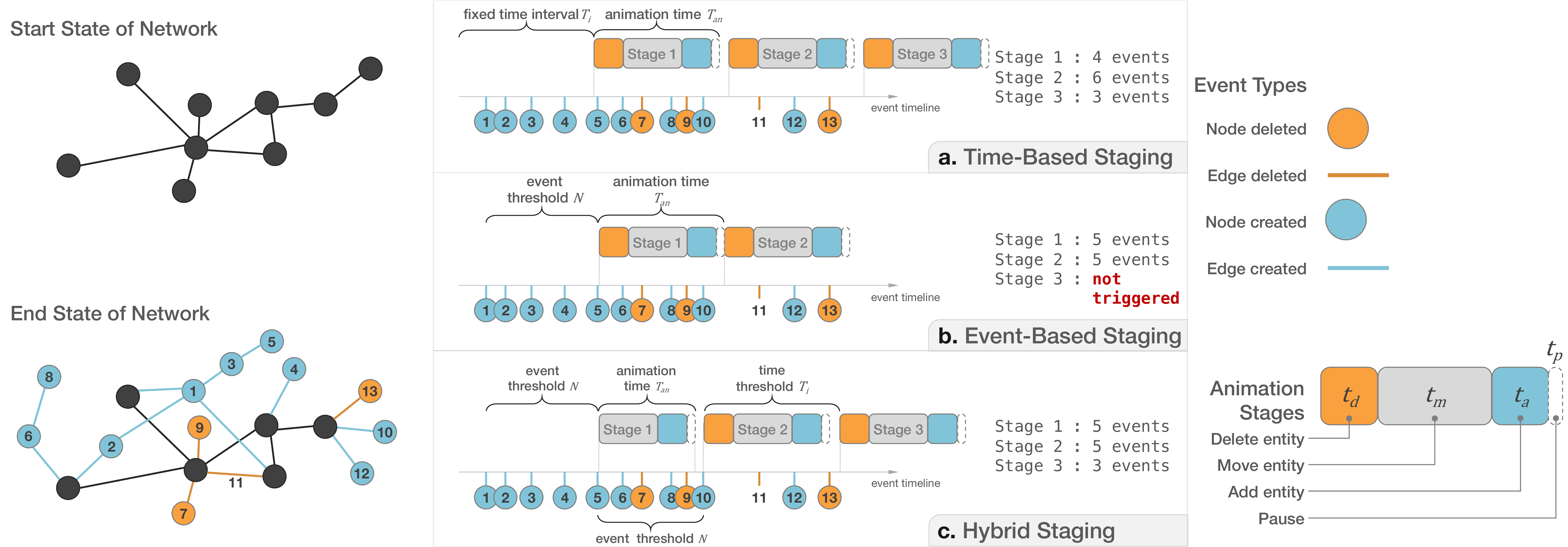}
  \vspace{-2em}
  \caption{Staging strategies shown for a sample dynamic network from an
    arbitrary start state (top left) to an arbitrary end state (bottom
    left).
    Nodes/edges that are created between these two states are marked in
    blue in the end state, while nodes/edges that are deleted are marked
    in orange.
    Each creation/deletion event is numbered based on their order of
    occurrence.
    The three animation staging strategies studied are shown in the
    middle: \textbf{(a)} time-based, \textbf{(b)} event-based, and
    \textbf{(c)} hybrid.
    Each animation stage, composed of deletion, movement, and creation
    sub-stages is shown for each staging strategy relative to the event
    timeline.
  }
  \label{fig:staging_strategies}
\end{figure*}

We base our animation design itself on Bach et
al.~\cite{Bach2014graphdiaries}, who
report that while deletion of nodes/edges and addition of nodes/edges
may both trigger positional changes to the network, pre-computing these
and ordering them as deletion-movement-addition works best to preserve
the viewer's mental map.
Our animation time $T_{an}$ will thus also be split into a time $t_d$
over which entities are deleted, time $t_m$ over which they are moved,
and a time $t_a$ over which new entities are added, in that order (see
bottom right of Fig.~\ref{fig:staging_strategies}).
In addition, we include a ``pause time'' ($t_p$) after each animation to
perceptually distinguish one stage from another.
Without a pause, the user would perceive a continuous series of
animations instead of the intended stages.
Thus, the total animation time $T_{an} = t_d + t_m + t_a + t_p$ \edit{as seen in Fig.~\ref{fig:staging_strategies}.}
Based on this, and on the requirements above, we examine three
strategies of staged animations.
For our study, we set $t_d$ = 450ms, $t_m$ = 600ms, $t_a$=450ms, and $t_p$=500ms, for a total animation time of $T_{an}$ of 2 seconds.

\textbf{Time-based staging:}
As the name implies, this staging strategy uses a specified set of time
intervals $t_i \ge t_a$ over which events are recorded.
Events are separated into deletions and additions, and the new positions
of the nodes in the dynamic graph are computed.
At the end of this time interval, the animation renders deletions,
movement to new positions, and additions in sequence.
This process is repeated for successive time intervals $t_i$.
One of the main advantages is that regardless of the data influx, the
animation will lag behind the actual events by a maximum of $t_i + t_a$,
satisfying the \textit{timeliness} requirement (R1).
However, this strategy does not place an upper limit on the number of
events per animation stage.
For instance, Fig.~\ref{fig:staging_strategies}~(a) shows three stages of
animation, with 4, 6, and 3 events respectively if represented using
time-based staging.
For high data influx, the number of events per stage can grow to an
overwhelming proportion.
Since there is no way of predicting the data influx for online dynamic
network, the effectiveness of this staging strategy hinges on the
data influx rate and the choice of $t_i$.
The time-based staging strategy is thus more suitable for
monitoring rather than comprehension tasks, provided the animations are not overwhelming.
For the purpose of our study, we use 2 seconds as the value of $t_i$.

\textbf{Event-based staging:}
Given that human visual perception is limited in terms of the number of
independently-moving objects it can track, it is reasonable to explore an
animation staging strategy that counts events as they occur, and deploy
the events once a threshold of $N$ events has been reached.
The separation of events and the subsequent animation follows the same
process as time-based staging, but there are a few differences in the
staging of the animation.
The number of events shown per stage is limited, which helps with
comprehension and mental map preservation (R2).
However, even when the data influx rate is high, only $N$ events can be
shown for every time period $T_{an}$, the animation time.
If the number of events that occur over $T_{an}$ is greater than $N$, it
results in events ``piling up'' to be animated, resulting in 
an increasing time lag between the event and its animation.
On the other hand, when the data influx rate is very low, such as when the time taken for $N$ events to occur is much higher than $T_{an}$, those events
are not shown at all until the event threshold is reached.
Fig.~\ref{fig:staging_strategies}~(b) illustrates this issue.
While three events have occurred after Stage 2, they are not shown to the viewer as the event threshold is $N=5$ in this example.
Both these conditions illustrate how an event-based staging strategy  \textit{does not} satisfy the timeliness requirement (R1).
Event-based staging strategies are thus more suitable for comprehension
tasks rather than those involving monitoring.
In our study, we use $N=5$ for comprehension tasks and $N=3$ for monitoring tasks.

\textbf{Hybrid staging:}
In order to address the shortcomings of time-based and event-based
staging strategies, we introduce a combination of the two: a
\textit{hybrid} strategy that uses both event and time thresholds to
trigger the next stage of the animation.
Thus the animation triggers at a specified event count $N$ or at a
specified time interval $t_i$, whichever occurs earlier.
For high data influx, the staging is based on an event-based trigger,
prioritizing comprehension (R2) over timeliness.
For low data influx, the staging is based on a time-based trigger,
prioritizing timeliness (R1).
Comprehension is not compromised in this case as the data influx rate is
low.
\edit{Traditional event- and time-based animations use uniform timings, even when there are no addition or deletion events.
This uniformity helps reduce mental load on the user as they can anticipate events.
However, when no events occur, this can contribute to the animation ``lag''.
Since our goal with the hybrid staging was to reduce lag without compromising comprehension, we use a variable animation time based on the kinds of events recorded for each stage.}
Thus, if there are no deletion events in a stage, the animation time
reduces to $T_{an} = t_m + t_a + t_p$.
Alternately if there are no addition events, the animation time reduces
to $T_{an} = t_d + t_m + t_p$.
There can also be cases where the time threshold $t_i$ is reached with
no events occurring, and this can trigger a convergence of the graph
involving only movement.
In this case, the animation time is simply $T_{an} = t_m + t_p$.
This reduces the transition time overall, making more effective use of
the viewer's short-term memory (R3).
The hybrid staging strategy thus combines the advantages of both
time-based and event-based strategies, \textit{except} in the case of
high-throughput data, where its timeliness is equivalent to that of
the event-based strategy.
However, the introduction of variable animation time reduces unnecessary
time delays even in this situation.
For our study, we maintain the same event and time thresholds for the hybrid staging as we do for event- and time-based staging.

We implement all three approaches for our user study in order to compare
these three strategies for both monitoring and comprehension tasks in
online dynamic networks.


\section{Implementation}

Given the focus of our study on monitoring and comprehension, our
implementation of staged animation of online dynamic graphs
precludes any user interaction with the system.
This is an artificial constraint as most visualization or visual
analytic systems for monitoring networks would have some control for
overview, filtering and playback.
Yet, it was necessary for our controlled study to exclude other
variables that might influence the perception of the results from each
staging strategy.
Our implementation thus focused on only one aspect, graph layout, other than the staging strategies.

A crucial aspect of dynamic graph layouts is how each time step is laid
out.
In offline dynamic graph layouts, a trade off is made between globally
optimized layout for all times or local optimized for each time.
Since online dynamic graph layouts do not posses the knowledge of all the
time steps ahead of time they focus on reducing the amount of movement
nodes experience between each time point. 
We utilize the incremental layout method work by Crnovrsanin et
al.\cite{Crnovrsanin2015}.
Their qualitative study demonstrated improved results over the two other state-of-the-art  methods: Aging~\cite{Gorochowski2012} and Pinning~\cite{Frishman2008}.
Their methods consist of two parts, an initial placement and layout
algorithm followed by a refinement approach.
The refinement allows nodes with high energy to move until they reach a
low energy state.
Due to how staged animation works and for the study, we had to make a few
changes to how their layout method operates.
In their implementation, refinement is run between time steps while the
system waits for more data.

\begin{figure}[t]
  \centering
  \includegraphics[width=\columnwidth]{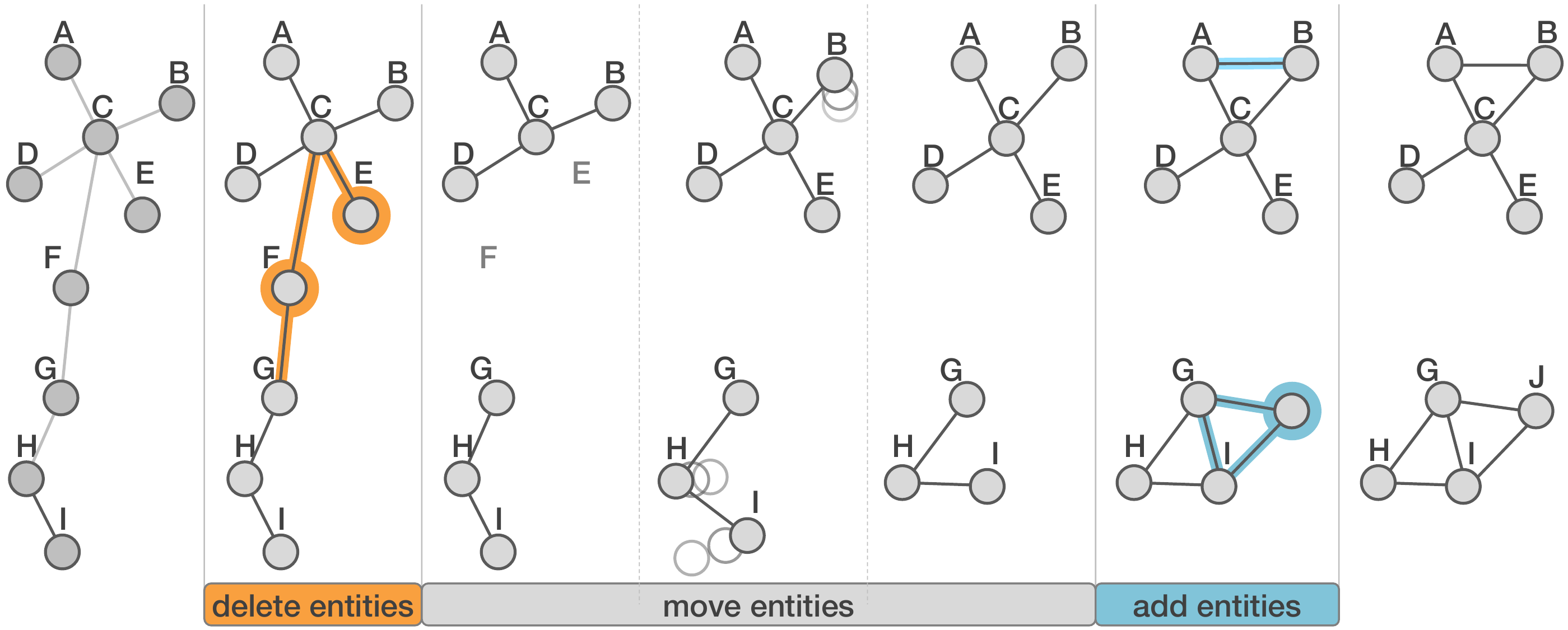}
  \caption{
    Illustration of how the staged animation appears to the
    participants in the study, inspired by
    GraphDiaries~\cite{Bach2014graphdiaries}.
    At the start of the deletion stage, all entities (nodes/edges) to be
    deleted flash orange, and then disappear over the next 0.5 seconds.
    Labels of deleted nodes remain briefly for the participants to
    register what nodes have disappeared.
    The remaining nodes move (1.2 sec) to their new computed positions.
    Entities that are added flash blue briefly and the color fades away
    (0.5 sec).
  }
  \label{fig:animation_order}
\end{figure}

The benefit of this approach is that it allows a gradual change of the
network and helps to maintain the stability of the network over time.
The incremental layout method that we use~\cite{Crnovrsanin2015} has two options: either refine the layout gradually between the time steps, or perform refinement and lay out the graph without showing the intermediate steps.
Unfortunately, in our implementation, we can not run refinement between
time steps due to constant data coming in the study.
Therefore, we run refinement right after the initial placement of new
nodes and edges, followed by the layout algorithm but before staged
animation is run. 
This groups the movement from both the addition and deletions as well as
high energy nodes shifting to a low energy state.
Another change is an addition of a central force for all nodes to keep
them closer to each other.
Without this central force, disconnected components would continue to
move away over time.
A side benefit is this allows us to fit all nodes within the screen,
making it easier to conduct the study. 
The actual disappearance, movement, and appearance of the entities in
each stage follows GraphDiaries~\cite{Bach2014graphdiaries} closely (see
Fig.~\ref{fig:animation_order}).

\section{User Study}
Our goal through the user study is to determine the suitability of each
staging strategy to a number of tasks broadly
based on task taxonomies for network visualizations with temporal
components~\cite{Andrienko2006exploratory, Ahn2013task,
Kerracher2015task}.
We broadly split our tasks into \textit{monitoring} and
\textit{comprehension} tasks, or elementary and synoptic tasks
respectively, according to the Andrienko Task Format
(ATF)~\cite{Andrienko2006exploratory}.
We designed a within-subjects study where each participant was exposed
to multiple monitoring and comprehension tasks using all three animation
strategies.
We supplement this study with a follow-up qualitative study with two experts in network visualization (see \autoref{sec:follow_up}).
We use a think-aloud protocol to evaluate their comprehension of a set of online dynamic network animation videos across the three animation strategies.

\subsection{Participants}
We recruited 21 participants (9 female, 12 male) between 18 and 35 years
of age.
All participants were university students, with 12 Ph.D students, 6
masters students, and 3 undergraduate students.
Eighteen students were computer science majors, the remaining
three majored in Aerospace engineering, Telecommunication, and
Information Technology, respectively.
Most of the participants (13 students) reported being highly familiar
with information visualization.
Five were moderately familiar with visualization, while three had little
to no familiarity with the subject.
Eight of the 21 participants reported using node-link diagrams regularly
in their work, an equal number reported having used them at least once,
while the remaining 5 had little or no knowledge of node-link diagrams.

\subsection{Apparatus}
Of the 21 participants, 16 used a MacBook Pro laptop with a 2.7 GHz
Intel Core i5 processor and 8 GB RAM, connected to a 30 in display
(2560$\times$1600 resolution).
Due to logistical constraints, five participants participated remotely,
and used different monitors (all 15-inch laptop screens).
All animations were shown to participants as video clips with no
playback controls.
The questionnaire including video playback was administered on a Chrome
browser.

\begin{figure}[t]
  \centering
  \includegraphics[width=\linewidth]{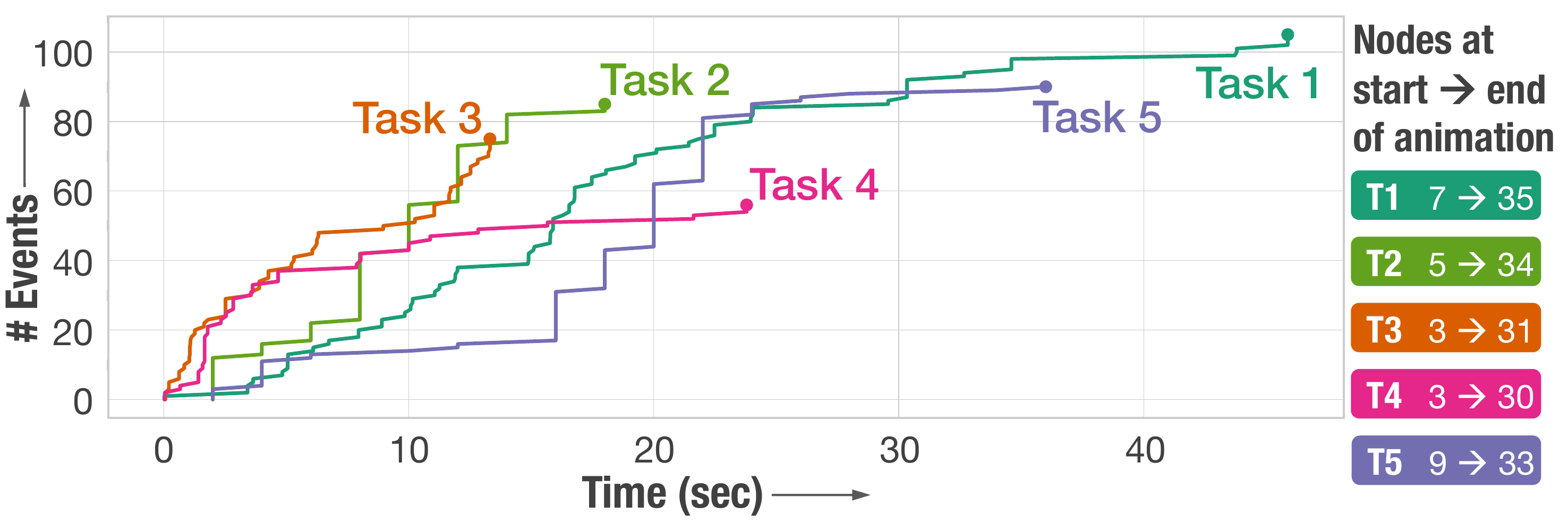}
  \vspace*{-8mm}
  \caption{
  Chart showing the rate at which the events occur over time in the raw data (left), \edit{and the number of nodes at the start and end of the animation for each task (right)}.
  The duration over which these events are displayed in the animations will depend on the animation strategy used.
  }
  \vspace*{-2mm}
  \label{fig:eventsvsTime}
\end{figure}

\subsection{Tasks \& Dataset}
\label{sec:tasks_dataset}

In order to generate the animations illustrating the three staging
strategies, we used the MIT Reality Mining
Dataset~\cite{Eagle2006reality}.
The dataset contains activity records of 100 individuals at MIT over the span of the 2004-2005 academic year and includes datasets recording proximity, location, communication, and other activity.
In this study, we draw from subsets of this data concerning the activities labelled \emph{Call} and \emph{Proximity}.
The \emph{Call} dataset contains temporal data of individuals placing calls to others, while the \emph{Proximity} dataset contains temporal data of individuals moving in and out of each other's Bluetooth ranges.
For our tasks, we choose temporal data segments from different sections of the larger dataset.
This allows us to choose segments with variations in the rate at which events occurred over time (see \autoref{fig:eventsvsTime}).
In addition, the two types of networks present in the data set (call and proximity) are structurally different from each other.

As mentioned earlier, tasks were split into monitoring and comprehension
tasks.
Monitoring tasks were timed.
The question was asked before the video was shown to the participants, and they needed to respond \textit{during} the video playback as soon as they spotted the answer to their question.
\edit{Participant response times were noted relative to when the event occurred, in addition to when the event was shown in the animation (see \autoref{fig:monitoring_results} for details).}
Comprehension tasks were multiple-choice, and required participants to
observe a video and then answer one or more questions related to the
video.
\edit{We used node labels for tasks that required paying attention to specific nodes, and kept the nodes unlabeled for tasks that were more general, e.g.\ paying attention to clusters or the overall graph.}
\autoref{tab:task_categorization} describes the tasks used in the study.
\edit{A third comprehension task was initially included in the study but subsequently dropped from the analysis due to the fact that the event-based and hybrid staging resulted in identical ``batches'' of animations, with only the node positioning being different. This would have created an unintended confound.}

Under Koussoulakou and Kraak's classification of spatio-temporal tasks~\cite{Koussoulakou1992spatia}, the monitoring tasks can be categorized as \emph{elementary}
under space and \emph{intermediate} under time, as participants are typically tracking one or two nodes over a given duration, looking for a specific behavior.
The comprehension tasks in this study as well as the follow-up think-aloud study fall under \emph{``overall level''} under both space and time in the same classification, as participants are asked to report on the overall behavior of a node, group of nodes, or the entire network over a time duration.
In terms of tasks specific to network visualization, we use as reference Ahn et al.'s taxonomy~\cite{Ahn2013task}, where temporal features are broadly classified into \emph{individual temporal features} that are typically event-related, \emph{shape of changes} that concern event collections such as growth/contraction, stability etc.\, and \emph{rate of changes} that involve the measurement of speed or time.
Since ours is an exercise in perception and not measurement, our tasks do not fall under the \emph{rate of changes} category.
Using this classification, our monitoring tasks can be categorized under \emph{individual temporal features}, while the comprehension tasks and the think-aloud tasks in the follow-up study can be categorized under \emph{shape of changes}.
Participants were given training questions for each type of task: one
question for monitoring, and two for comprehension.

\begin{table}[h]
  \small
  \centering
  \begin{tabular}{l c p{2.1in} }
    \toprule
    \textbf{Task Type} & \textbf{ID} & \textbf{Task Description} \\
    \midrule

    \multirow{3}{*}{Monitoring}
      & T1 & Track clusters and respond as soon as they merge. \\
      & T2 & Track graph and respond as soon as a particular named
             entity (node) appears. \\
      & T3 & Track graph and respond as soon as two named entities
             (nodes) are linked. \\
    \midrule
    \multirow{7}{*}{~\newline ~\newline Comprehension}
      & T4 & Entity pointed out before the video, and after the video
             playback, asked what happened to it over the course of the
             animation.\\
      & T5 & Cluster pointed out before the video, and after the video
             playback, asked what happened to it over the course of the
             animation. \\
    \bottomrule
  \end{tabular}
  \vspace{1mm}
  \caption{Task categorization and description for each staging
  \label{tab:task_categorization}
  strategy}
  \vspace{-3mm}
\end{table}

\edit{In real-world applications such as our security networks example, the individuals who monitor the networks are intimately familiar with said network.
Based their knowledge of prior attacks, they can judge which nodes are vulnerable and need attention.
While it would be difficult to (a) find a suitable number of network security experts and (b) set up the data to suit their prior knowledge of similar networks, we were able to \textit{simulate} this prior knowledge by asking participants to pay attention to certain nodes.}

\edit{
In addition, real-world scenarios would likely have additional embellishments such as highlighting on recently-changed portions of the network visualization.
However, we decided to use it sparingly for a number of reasons.
Given our focus on understanding which staging strategies best use participants' capabilities to discern changes in networks, we use highlighting as shown in Fig.~\ref{fig:animation_order} to draw attention to node creation and removal.
Other forms of highlighting that draw attention to certain categories of behavior often do so for the user to step back in time and review what happened to the highlighted nodes.
This would make sense in a long-term case study of an actual network being monitored, but not in our controlled study.}


Since most participants were not expected to be familiar with dynamic
graphs and node-link diagrams, the questions were mostly phrased in the
context of a social network.
For instance, T3 was worded as \textit{``You will be shown a friendship
network with users shown as nodes and relationships between them shown
as edges. Look out for Ryan and Emily, and click on the button as soon
as they become friends.''}
This being a full-factorial within-subjects design, tasks \edit{T1--T5} were
repeated for each condition, i.e. each animation staging strategy.
We used the same dataset for each condition, with labels changed and
graphs rotated/mirrored for each condition.
While the graphs were all identical (except for rotation/mirroring) at
the start of the animation, the different staging strategies would
result in changes in layout between the three staging conditions.
While this potentially introduces an additional variable (layout) into
our study, it is unavoidable within the scope of this work.
All tasks were performed in sequence  \edit{(T1--T5)} for each condition.
We counterbalanced the condition order using a Latin Square design to mitigate learning effects.

\subsection{Procedure}

Individual participants were first given a basic background of the
study, but were not described the specific animation techniques, so as
to not bias them.
Instead, they were told that data would be shown to them in three
different forms, and that they would be asked to perform a set of tasks
for each form of data presentation.
All questions and tasks were presented in the form of an online survey,
with the video showing animations embedded in the survey.
To ensure participants see the video only once, playback controls were
disabled.
Due to the animation strategies used, the video durations varied for
each condition. 
Table~\ref{tab:video_durations} shows video duration for each task and condition.

\begin{table}[h]
  \small
  \centering
  \begin{tabular}{c c c c }
    \toprule
    \multirow{2}{*}{\textbf{Task ID}} &
    \multicolumn{3}{c}{\textbf{Video duration (sec)}}\\
    \cmidrule{2-4}
     &
    \textbf{Event-based} &
    \textbf{Hybrid } &
    \textbf{Time-based} \\
    \midrule
    T1 & ~34 & 34 & 16 \\
    T2 & ~75 & 65 & 23 \\
    T3 & 108 & 96 & 15 \\
    T4 & ~69 & ~65 & ~28 \\
    T5 & ~92 & ~75 & ~37 \\
    \bottomrule
  \end{tabular}
  \vspace{1mm}
  \caption{
    Video durations for each task and staging strategy.
    Note that the data shown is the same for each task; the varying
    durations are a result of the staging strategies.
    Note that for T1--T3 (monitoring), participants are not required to
    watch the entire video.
  }
  \label{tab:video_durations}
  \vspace*{-5mm}
\end{table}

In the case of monitoring tasks, the video was accompanied by a button
labelled \textit{``Click as soon as you find (the answer)''}, and the
time elapsed between the start of video playback and the button click
was recorded.
Note that for monitoring tasks, participants do not have to watch the
entire video.
Participants were presented with all 5 tasks in the same order for each
condition (animated staging strategy).
Our study thus involved 21 participants $\times$ 3 staging strategies
$\times$ 5 tasks, resulting in a total of 315 trials.
At the end of each condition, participants also filled out a NASA Task
Load Index (TLX) response sheet~\cite{Hart1988development}.
Note that we collected the NASA TLX data once per condition rather than
once per task to avoid survey fatigue.
A typical session lasted 45 minutes.

\subsection{Hypotheses}
%

Based on our design considerations and the requirements outlined in
Section~\ref{sec:design_considerations}, we formulated the following
hypotheses:
\vspace{-2mm}

\begin{enumerate}[itemsep=-0.5mm]
  \item[\textbf{H1}] Participant response to the monitoring tasks will
     \edit{be affected by the volume of data. They will} be quicker for event-based staging conditions than the remaining two as the reduced visual complexity would help them spot the event as soon as it happens.
    For the same reason, we posit that hybrid staging will prompt
    quicker responses than time-based staging.
    \edit{Response time is a stand-in for the ability to perceive an event that occurred.}
  \item[\textbf{H2a}] Participant responses to comprehension tasks will
    show more errors in time-based staging conditions than the remaining
    two due to the increased visual complexity to which time-based
    staging is susceptible given a high rate of data influx.
  \item[\textbf{H2b}] Participant responses to comprehension tasks will
    show more errors in event-based staging conditions than in hybrid
    staging conditions due to the greater time period for which
    participants need to track and remember events.
  \item[\textbf{H3a}] Participants will report lower levels of
    performance and higher levels of frustration in time-based staging
    conditions when compared to the remaining two.
  \item[\textbf{H3b}] Participants will report higher levels of mental,
    physical, and temporal load and effort in time-based staging
    strategies when compared to the remaining two.
\end{enumerate}

\vspace{-3mm}

\section{Results}
We split our results into monitoring (hypothesis H1),
comprehension (H2a \& H2b), and participant experience (H3a \& H3b), and
report the results in detail under each.

\begin{figure}[tb]
    \centering
  \includegraphics[width=0.8\columnwidth]{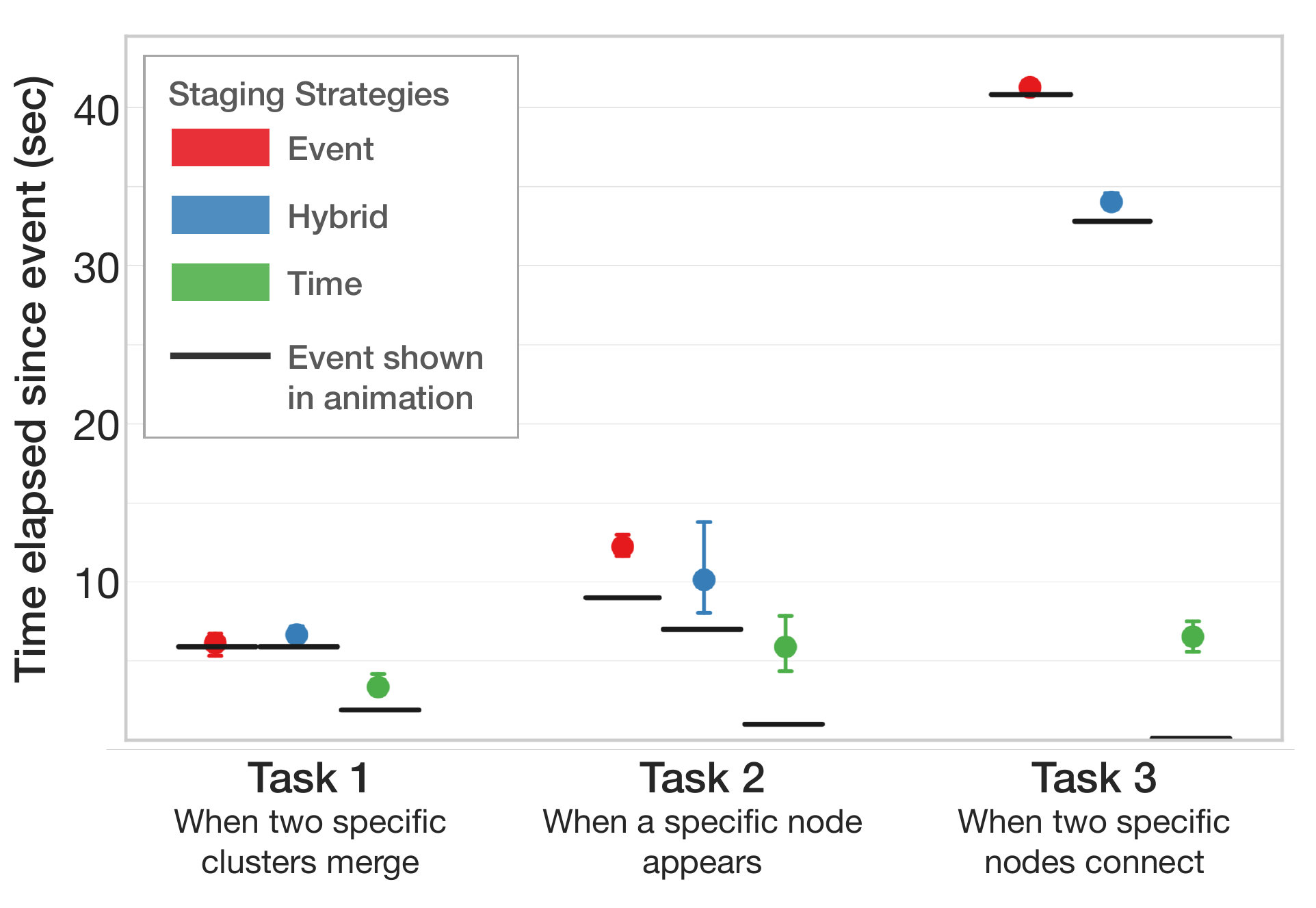}
  \vspace{-3mm}
  \caption{
    Differences between \edit{the time at which an event occurs}
    (\edit{i.e.\ the \textit{y=0} line}),
    the time at which the event is shown in the animation (black horizontal
    lines), and participant response times \edit{(colored markers with error bars)} for monitoring tasks (T1--T3).
    Error bars represent 95\% CI.
  }
  \label{fig:monitoring_results}
\end{figure}

\subsection{Monitoring Tasks}
\begin{figure}[t]
    \centering
  \includegraphics[width=.8\columnwidth]{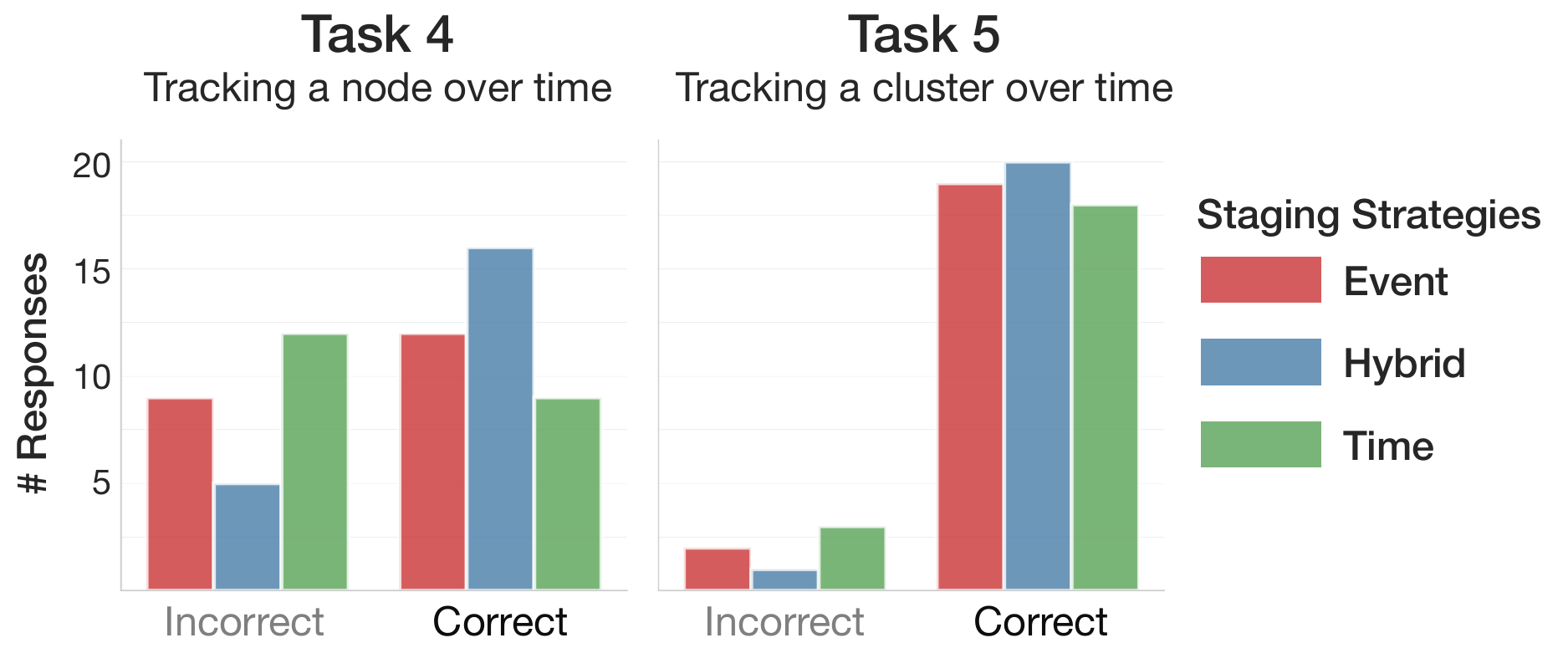}
  \vspace{-3mm}
  \caption{
    Distribution of correct and incorrect answers for the comprehension
    tasks (T4 \& T5), categorized by animation staging strategy.
  }
  \label{fig:comprehension_results}
  \vspace{-3mm}
\end{figure}

Fig.\ref{fig:monitoring_results} shows participant response distribution
for monitoring tasks (T1--T3).
The figure shows two kinds of delays: (1) the delay in participant
response when compared to the actual event (the boxplots in the figure)
and (2) the delay between an event occurring and it being shown in the
video (the horizontal \edit{black} lines in the figure).
We separate the two delays.
We measure the difference between the time at which the participant
is shown the event and their corresponding response as the ``participant
response time''.
We use this as a measure of how easily the participant was able to see
the event when it was shown to occur (hypothesis H1).
\edit{This also addresses cases where participants respond \textit{before} the event occurs, which in monitoring tasks is an error at the same level or worse than a \textit{delayed} response.
We observe in the raw data that this happened only for three participants in time-based staging.}

We analyzed the participant response times using a
repeated-measures analysis of variance (RM-ANOVA) and found a
significant effect of animation staging strategy on the time difference
between participant response time and the time at which the event was
shown on the video for tasks T1 ($F(2, 40) = 4.04, p<0.05$)
and T3 ($F(2, 40) = 99.43, p<0.001$).
No significant difference in response times was found between the
conditions for task T2.
This partially confirms hypothesis H1.
A post-hoc Tukey HSD test showed significant pairwise differences
between time-based and hybrid staging conditions ($p<0.05$) for tasks T1
and T3, and between time-based and event-based staging conditions
($p<0.001$) for task T3 (see \autoref{tab:monitoring_tukey_results}).

\begin{table}[tb]
  \small
  \centering
  \begin{tabular}{c l c c c c c}
    \toprule
    \multirow{2}{*}{Task} & \multirow{2}{*}{Condition} & 
    \multicolumn{2}{c}{Response Diff. (Video)} &
    \multicolumn{3}{c}{Tukey HSD Significance} \\
    \cmidrule(lr){3-4}
    \cmidrule(lr){5-7}
     & & Mean (sec) & S.D. (sec) & Event & Hybrid & Time \\
     \midrule
        & Event  & 0.233 & 1.580 & --         &    & \textbf{*} \\ 
     T1 & Hybrid & 0.753 & 1.158 &            & -- &            \\ 
        & Time   & 1.454 & 1.606 & \textbf{*} &    & --         \\ 
     \cmidrule(lr){1-7}
        & Event  & 3.230 & 1.522 & --      &         &         \\ 
     T2 & Hybrid & 3.132 & 7.589 &         & --      &         \\ 
        & Time   & 4.892 & 4.046 &         &         & --      \\ 
     \cmidrule(lr){1-7}
        & Event  & 0.464 & 0.788 & --      &         & \textbf{**} \\ 
     T3 & Hybrid & 1.213 & 1.164 &         & --      & \textbf{**} \\ 
        & Time   & 6.722 & 2.294 & \textbf{**} & \textbf{**} & -- \\ 
     \midrule
     & & & \multicolumn{2}{c}{\textbf{*} : $p<0.05$} &
     \multicolumn{2}{c}{\textbf{**} : $p<0.001$} \\
     \bottomrule
   \end{tabular}
   \vspace{1mm}
   \caption{Results of monitoring tasks, with pairwise significant
   differences between conditions.}
   \label{tab:monitoring_tukey_results}
   \vspace{-2mm}
 \end{table}

%
\subsection{Comprehension Tasks}

\autoref{fig:comprehension_results} shows the distribution of correct
and incorrect answers for all the comprehension tasks, categorized by
the staging conditions.
We analyzed the distribution of correct and incorrect responses for each
question using Cochran's Q-Test.
Overall, we found no significant difference in response correctness
between conditions for tasks T4 and T5 (rejecting H2a and H2b).

\subsection{Participant Experience}

\begin{figure}[t]
  \centering
  \includegraphics[width=\columnwidth]{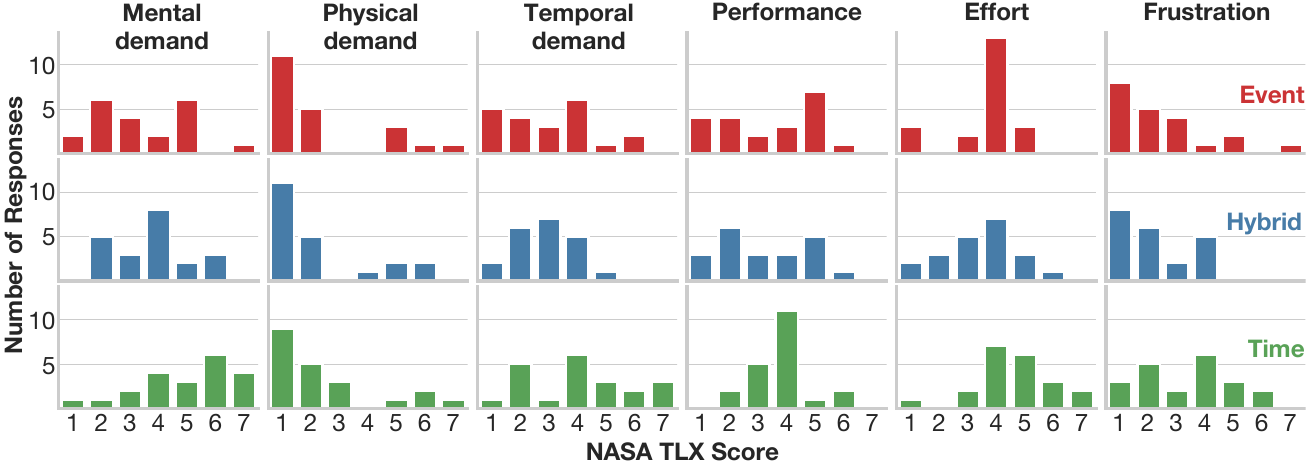}
  \vspace*{-6mm}
  \caption{
    Distribution of participant responses showing the mental, physical, and temporal
    demands of the tasks, along with their perception of performance,
    effort, and frustration while doing the tasks.
    Scores were self-reported on the 21-point NASA TLX
    scale and converted to a 7-point scale for comprehension.
  }
  \label{fig:tlx_results}
\end{figure}

The distribution of participant responses on the NASA TLX
for each staging condition is shown in Fig.~\ref{fig:tlx_results}.
We performed Friedman's test on each scale separately, and found
significant differences between the staging conditions for
participant responses on 
mental demand ($\chi^2=15.89, p<0.001$),
physical demand ($\chi^2=8.79, p<0.05$),
temporal demand ($\chi^2=9.66, p<0.01$),
effort ($\chi^2=5.05, p<0.001$), and
frustration ($\chi^2=11.14, p<0.01$).
No significant difference was found 
for participant responses on performance.
A posthoc Conover test revealed pairwise significant
differences(see Table~\ref{tab:tlx_friedman_results}).

\begin{table}[tb]
  \small
  \centering
  \vspace{-5mm}
  \begin{tabular}{l l r r r r r}
    \toprule
    \multirow{2}{*}{Task} &
    \multirow{2}{*}{Condition} & 
    Score &
    \multicolumn{3}{c}{Conover Test Significance} \\
    \cmidrule(lr){4-6}
       & & (median) & Event & Hybrid & Time \\
     \midrule
            & Event  & 9  & --          &             & \textbf{**} \\
     Mental & Hybrid & 10 &             & --          & \textbf{**} \\
            & Time   & 15 & \textbf{**} & \textbf{**} & --          \\
     \cmidrule(lr){1-6}
              & Event  &  3 & --          & \textbf{*} & \textbf{**}\\
     Physical & Hybrid &  3 & \textbf{*} & --          & \textbf{**}\\
              & Time   &  4 & \textbf{**} & \textbf{**} & --         \\
     \cmidrule(lr){1-6}
              & Event  &  9 & --          &             & \textbf{**} \\ 
     Temporal & Hybrid &  7 &             & --          & \textbf{**} \\ 
              & Time   & 11 & \textbf{**} & \textbf{**} & --          \\ 
     \cmidrule(lr){1-6}
              & Event  & 10  & --          & \textbf{**} & \textbf{**}\\
     Effort   & Hybrid & 10  & \textbf{**} & --          & \textbf{**}\\
              & Time   & 13  & \textbf{**} & \textbf{**} & --         \\
     \cmidrule(lr){1-6}
                 & Event  &  5 & --          & \textbf{**} & \textbf{**}\\ 
     Frustration & Hybrid &  6 & \textbf{**} & --          & \textbf{**}\\ 
                 & Time   & 11 & \textbf{**} & \textbf{**} & --         \\
     \midrule
      & & \multicolumn{2}{c}{\textbf{*} : $p<0.05$} &
     \multicolumn{2}{c}{\textbf{**} : $p<0.01$} \\
     \bottomrule
   \end{tabular}
   \vspace{1mm}
   \caption{Participant responses on the NASA Task Load Index for each
   condition with pairwise significant differences marked.}
   \label{tab:tlx_friedman_results}
   \vspace{-7mm}
 \end{table}

From the table, we see that participants were significantly more
frustrated using time-based staging than the other two,
though participant self-report of performance showed no
significant differences (partially confirming H3a).
On the other hand, mental load, physical load, temporal load, and effort
were all significantly higher for time-based staging than for the
remaining two (confirming H3b).
In fact, with the exception of mental load, there are significant
differences in participant perception of load, effort, and frustration
between hybrid and event-based strategies as well, with hybrid
performing better than event in terms of temporal load and worse in the
other measures.

\section{Follow-Up Study with Experts}
\label{sec:follow_up}

The participants in the prior study did not have much expertise in network visualization, though they had varying degrees of familiarity with the subject.
To follow up our findings with observations from expert participants, we conducted a qualitative study of the three animation strategies with domain experts in network theory and visualization.
The participants (P1 and P2) were Graduate Ph.D.\ students with over 5 years of experience in network analysis and visualization, and they design and implement new
network visualization techniques. 

We followed a think-aloud protocol where each participant was shown an animation of a dynamic network, and asked to narrate aloud what they thought was happening throughout the animation.
At the end of each animation, they were asked follow-up questions on what they had just observed, and about the state of the network in general, and observations they had made in particular.
The video was played only once (during the think-aloud component), and participants were asked not to play back the video.
The study was administered remotely via web conference, with both participants using Mac Book Pro laptops with resolutions of 2560$\times$1600 and 1920$\times$1080 each.

The study involved 3 different animation clips for each animation strategy.
One clip used the MIT proximity dataset~\cite{Kent2015comprehensive} mentioned in \autoref{sec:tasks_dataset},
while the remaining two clips used the LANL dataset mentioned in the introduction.
For the MIT dataset, participants were asked to observe and track individuals that moved through the network the most, and individuals who had high centrality, i.e.\ those who connected two or more clusters in the network.
For the LANL dataset, participants were asked to keep track of computers that connected to multiple other computers, and those that switched connections between different computers frequently.
Participants viewed training clips to familiarize them with the kind of data they would observe and their contexts.
They were shown video snippets describing what behaviors in the network we are trying to identify.
For example, we showed the users video snippets of what a stable network would be.
As in the previous study, the same data was animated using the three different strategies and shown to the participants.
Learning effects were minimized by (a) ordering the clips so that the same data was not shown in successive clips, and (b) changing the node identifiers between animation strategies.
Finally, the order in which the videos were shown were switched up between participants.
Our observations are described below.

\textbf{Time-Based Animation.}
Participants' responses to time-based animation clips seemed rushed.
They were unable to follow sudden changes that occurred in the network, and found it difficult to track individual node connections.
To catch up to the rapid changes in the network, the participants went from using specific network terminologies to describe changes to general phrases such as ``\textit{I see a lot of changes in the network}" (P1), and ``\textit{changes happening everywhere}" (P2).
However, we noticed that both participants provided similar descriptions of the overall network evolution, especially attributes such as changes in network size and stability of clusters.

\textbf{Event-Based Animation.}
Participants found certain sections of the network to be quite dense, stating that they seemed like a ``hairball" which made it difficult to identify which nodes were connected to which others.
The participants were able to identify the overall network evolution trend which included descriptions of stability and cluster changes.
There were sections of the video where both the participants claimed to seeing ``\textit{not many changes}".

\textbf{Hybrid Animation.}
Participants were able to identify, with little effort, the stable and unstable sections of the network, the nodes that connected different parts of the network together, and the overall network evolution trend.
Their description of the network evolution was less rushed.
The participants mentioned that the dense sections of the network were ``\textit{hard to follow and explain}". However, they were still able to explain individual connectivity within the dense network which wasn't the case in event and hybrid tasks.

\section{Scalability}
\label{sec:scalability}

\begin{figure}[t]
  \centering
  \includegraphics[width=0.95\linewidth]{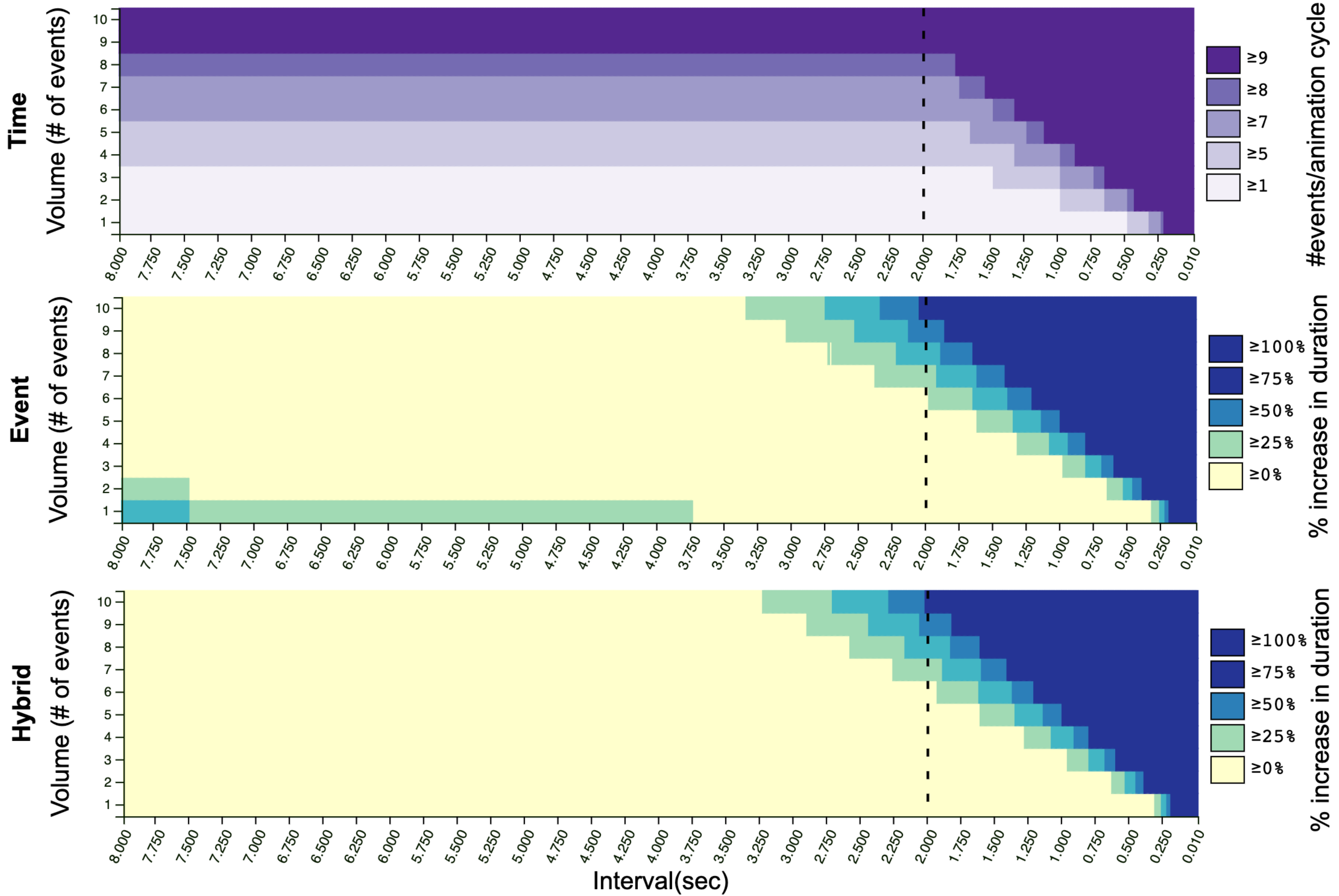}
  \vspace{-1em}
  \caption{
  Results of simulating individual animation strategies for varying event occurrence rates and time intervals.
  }
  \vspace{-4mm}
  \label{fig:scalability}
\end{figure}

We explore how scalability affects each of the animation strategies by varying the volume of data (i.e., number of events) and the time interval in which they occur.
Testing user perception of a wide range of event counts and intervals with a user study would be unrealistic due to the large number of possible variables. 
Instead, we ran a simulation of each animation strategy for one minute. 
For the time-based animation strategy, all events are shown in near real time.
The limiting factor is the number of events a user can perceive at a given time.
Therefore, we output the number of events shown per each animation cycle. 
As previously stated, studies haven shown that individuals can successfully track up to five simultaneously moving objects~\cite{Pylyshyn1988tracking}. 
Based on this constraint, we limit the number of events that can be shown at one time to 5 for event-based and hybrid strategies.
This constraint will affect the delay in seconds between the time an event occurs and the time that it is shown in the animation.
For the event-based strategies, we also look into the ``offset''---the time taken for 5 events to accumulate before being animated---which would depend on the data influx rate.

We vary the volume of data (i.e., the number of events) from 1 to 10 events at a time, and vary the time interval over which these chunks of events occur from 8 seconds down to 0.001s.
The results of our simulation are shown in Figure \ref{fig:scalability}.
The dashed line at the 2 second mark represents the point when the rate at which events occur is greater than the rate at which they are shown in the animation.
The number of events that occur is mapped to the y-axis, the time taken for each interval is mapped to the x-axis, and color is mapped to either the time delay (for event-based and hybrid animations) or number of events displayed per animation cycle (for time-based animations).

We see a staircase effect for the time-based strategy results (\autoref{fig:scalability}).
Intervals longer than 2 seconds follow a predictable pattern of being dependent on the number of events 
in that interval as there is enough time for the animation cycle to complete before new events occur. 
With an increase in the number of events occurring per interval, the ability to perceive the events quickly declines as the events exceed 5 per interval.

The simulation results for event-based and hybrid strategies in \autoref{fig:scalability} show a similar staircase effect that extends beyond the animation cycle of 2 seconds.
Since there is a cap of 5 events that can occur per any animation cycle, multiple cycles are needed to handle a high volume of data.
In either case, a pile-up of events occurs to the point that the animation cycles can not keep up.
For instance, 10 events occurring every two second interval means
that this set of events will take up the next two animation intervals, during which time, 20 more events have accumulated.
The event-based strategy shows an offset at the bottom when fewer than five events occur over a standard animation cycle.
That cycle will not be triggered until five events have accumulated, thus increasing the time delay between the first event occurring and it being shown.
This simulation confirms our notion that the hybrid strategy eliminates the inordinate delays that we see in event-based animations with low event rates, and works in a manner similar to event-based strategy at higher event rates.
It also indicates that the hybrid strategy's timeliness is not better than event-based strategy for high event rates.
\vspace{-2mm}

\section{Discussion}
We will first summarize the results from our analyses before explaining
and generalizing them.
Of our proposed hypotheses, we find:
\vspace{-2mm}

\begin{itemize}[label=$\bullet$,leftmargin=1.5em]
\setlength{\topsep}{-0.08in}
\setlength{\itemsep}{-0.04in}
  \item Event-based staging showed  significantly shorter
    response times compared to time-based staging, but not compared to
    hybrid staging for two of the three monitoring tasks.
    Hybrid staging showed shorter response times
    than time-based staging for one monitoring task (\textbf{partially
    confirming H1})
  \item Participant responses to comprehension tasks showed no
    significant differences for tasks T4 and T5.
    These results \textbf{reject H2a} and \textbf{H2b}.
  \item Participant responses to the NASA TLX showed no significant
    difference in participant perception of performance, but a
    significantly higher level of frustration in time-based staging
    compared to the remaining conditions, and in hybrid staging
    compared to event-based staging (\textbf{partially confirming
    H3a}).
  \item Participants also reported significantly higher mental,
    physical and temporal loads and effort on time-based staging
    compared to the remaining two (\textbf{confirming H3b}).
    They also reported higher physical load and effort in event-based
    staging compared to hybrid staging.
\end{itemize}
      
\subsection{Explaining the Results}
We had posited that timeliness (R1), mental map preservation (R2), and
minimization of transition time (R3) were the driving requirements in
staging animations in online dynamic networks.
Our hypotheses were derived from these requirements, and while they were partially confirmed for monitoring tasks, they were rejected for the comprehension tasks.
In this section, we examine the instances where the hypotheses failed and why they failed.

\subsubsection{Monitoring Tasks}
The three monitoring tasks were each designed to involve more complex
monitoring than the previous task, with fewer changes to the graph in
T1, and more and more complex changes over longer time durations (see
Table~\ref{tab:video_durations}).
In addition, T1 was an abstract task
with unlabeled nodes, while T2 and T3 involved graphs with nodes labeled
as first names of people.
Specifically, T2 simply required participants to look for the
appearance of one (named) node, while T3 required them to look for two
named nodes, track them, and respond when they connect.
It is very likely that the differences between staging conditions were
less significant for easier and ``familiar'' tasks (such as T2) while
they were more pronounced for more complex tasks (T3).
While differences between the hybrid and event-based strategies
do not emerge even for the complex monitoring task (T3), the answer
perhaps lies in the comprehension task results.

\subsubsection{Comprehension Tasks}
The argument of complexity can be made to explain the results of the
comprehension tasks as well.
Of the first two comprehension tasks, T4 appears to be too complex with
an almost equal distribution of right and wrong answers regardless of
the condition.
T5 appears to be too straightforward, with most participants answering
correctly regardless of condition.
T4 gives participants the label of a node to track, and requires them to
first seek out the node once the video starts (the node appears
\textit{after} the start of the video), and keep track of it,
remembering the changes in its degree.
Participants performed poorly regardless of the condition likely because
of their failure to identify the node in time, missing early changes to
the node's degree.
T5 demands the least from the participant's perception and
memory as the cluster is labeled and does not go through very complex
changes (it grows and splits).

While the follow-up study with expert participants was not quantitatively evaluated, it appears to validate our reasoning: participants could only make general observations about the network in time-based animations, such as changes in network size and cluster stability.
Participants observed that event-based animations sometimes had little or no changes occur over certain periods, while their impression of hybrid animations fell somewhere between the two, skewing towards event-based animation.
This observation is supported by the scalability simulation discussed in \autoref{sec:scalability}.
Given the limitation posed by the number of perceivable events~\cite{Pylyshyn1988tracking}, the hybrid strategy cannot be more ``timely'' than event-based strategy for high event occurrence rates.
Other strategies that use Gestalt principles of completeness and common fate need to be adopted to group related events together so that they are perceived as one event.
This can theoretically improve the users' perception of multiple events, though it may come at the risk of reduced perception of anomalous activity.

\subsubsection{Participant Experience}
Responses on the NASA TLX scale were as predicted, except for ``performance'', the mixed responses for which could be because participants were not informed whether they had the
correct answer.

\subsection{Generalizing the Results}
Overall, event-based tasks scored well on participant preference as they
reduced the load on participant perception (R2).
On the other hand, Fig.~\ref{fig:monitoring_results} clearly shows that
time-based strategies are best for timeliness (R1), which is achieved without compromising comprehension in the case of
low data influx.
The hybrid approach tries to bridge this gap between timeliness and
comprehension by providing both event-based and time-based thresholds.
An adaptive hybrid strategy that combines shorter time thresholds with a
higher event threshold or vice versa has the potential---with judicious
threshold choices---to provide the regularity of updates and the
timeliness of time-based transitions for low data influx, and the ease
of comprehension for high data influx.
Even in the case of a monitoring task, we see from the study that
comprehension is to be prioritized over timeliness. 

\edit{
It is worth noting that the general approach of binning---used for any staged animation---comes at the expense of information loss within the bins.
This includes event order within the bin, and entire events themselves---such as nodes appearing and disappearing within the binned intervals.
The limitations can be overcome by coupling additional views and metrics to track behaviors within a bin window and notify the user when such instances occur.
The effectiveness of these staging alone will vary on the volume and rate of the incoming data. 
With higher rates of data influx, an adaptive staging strategy that adapts to the \textit{complexity} of the changes.
For instance, a large change can still be simple if the change is of one kind, e.g.\ a cluster of nodes being added to the network.
Another approach would be to combine such an adaptive staging (for higher data influx rates) with time-based staging (for lower rates), to reduce the animation lag.
At any rate, animation alone is not sufficient to monitor and comprehend online dynamic networks.
Instead, it might work to the dashboard designer's advantage to prioritize comprehension when it comes to animation staging, and provide supporting views for monitoring tasks.}

\section{Limitations and Future Work}

Through our study we learned that regardless of monitoring or
comprehension tasks, animation staging strategies that prioritize
comprehension do better for participant response times, accuracy, and
comfort.
Yet, the differences between the staging strategies are slightly blurred
for tasks that are less complex or require less monitoring time.
In addition, our hybrid strategy was a simple combination of the
parameters used in the time-based and event-based strategies.

\edit{How event- and time-based parameters are combined for the hybrid strategy---and the user's awareness of the strategy---could impact monitoring tasks. In our study, participants were not informed of the delays between the actual event times and when they were shown in the animations (black bars in \autoref{fig:monitoring_results}). Their perception of their response time to the event was thus different from the actual response time to the event.
In addition, we used constant time/event thresholds for each animation strategy for our study.
We plan to explore the design space of adaptive strategies discussed in the previous section, along with indicators for animation lags in the future.
In addition, we plan to explore using these staging strategies to capture network ``states'' that will then be visualized as static, small multiples visualizations to be used for post-event analysis.
Lastly, we plan to incorporate the hybrid staging strategy into a visual analytic system for analyzing online dynamic networks to examine its applicability in real-world scenarios.
}

\acknowledgments{
This research is sponsored in part by the U.S. National Science Foundation through grants IIS-1741536 and IIS-1528203.} 

\bibliographystyle{abbrv-doi}

\bibliography{staged-animation-network}
\end{document}